\documentclass[amssymb,prl,aps,showpacs,twocolumn]{revtex4}

\usepackage{mathrsfs}
\usepackage{amsmath}

\begin{document}
\title{A new electromagnetic wave in a pair plasma}   

\author{Lennart Stenflo, Gert Brodin, Mattias Marklund, and Padma K. Shukla}
\affiliation{Department of Plasma Physics, 
Ume{\aa} University, SE-901 87 Ume{\aa }, Sweden}

\date{Received 26 April 2004}

\begin{abstract}
A new nonlinear electromagnetic wave mode in a plasma is reported. Its
existence depends on the interaction of an intense circularly polarized
electromagnetic wave with a plasma, where quantum electrodynamical 
photon--photon scattering is taken into account.  As an illustration, we 
consider a pair plasma and show that the new mode can be significant in 
astrophysical settings and in the next generation laser-plasma systems. 
\end{abstract}
\pacs{52.27.Fp, 52.35.Mw, 52.38,-r, 52.40.Db}

\maketitle

The concept of photon--photon scattering, an effect stemming from the
interaction of photons with virtual electron--positron pairs in
quantum electrodynamics (QED), has attracted attention in a number 
of recent publications (see, e.g. Refs.\
\cite{Shen-etal,Shen-Yu,Marklund-Brodin-Stenflo,Shukla-Eliasson} and 
references therein). Although a weak effect, it is believed to be an
active ingredient of highly magnetized neutron stars \cite{magnetar},
giving rise to frequency down-shifting \cite{Harding}
and lensing \cite{Shaviv-etal}. Moreover, it has been suggested that
the next generation laser-plasma systems would be able to produce
field strengths close to the Schwinger field $ \sim 10^{18}$ V/m
\cite{Bulanov-etal}, at which the vacuum becomes fully nonlinear, and
photon--photon scattering needs to be taken into account for a full
understanding of the laser-plasma dynamics.     

Lasers are currently reaching powers of
$10^{21}-10^{22}$ W/cm$^2$ \cite{mou98,taj02}, and an electron in such
fields can attain a highly relativistic quiver velocity.  
Moreover, a plasma under the influence of such intense lasers will be
highly nonlinear and the relativistic dynamics of the laser-plasma
system will thus be complicated, giving rise to, e.g.  
soliton formation and pulse collapse \cite{Shukla-etal}. 
It has been noted that the latter of these effects could be of
interest for generating electromagnetic field intensities above the
Schwinger limit \cite{cai04,Bulanov-etal,bob,bob2}.
Furthermore, laser pulse collapse and the subsequent electron
expulsion due to the immense ponderomotive force is able to
create plasma channels in which ultra-high intensity field strengths
are reached \cite{Yu-etal}, such that the  nonlinear vacuum effect of 
elastic photon--photon scattering could become important \cite{Shen-Yu}. 

In pulsar surroundings, pair production and pair plasmas play an
important role in the dynamics of these environments (see, e.g.
Refs.\ \cite{magnetosphere,Arendt-Eilek,Asseo}). Charged particles will
attain relativistic energies close to the pulsar magnetic poles and
radiate $\gamma$-ray photons. This, together with the super-strong 
magnetic field present around these objects \cite{magnetosphere},
is believed to produce a pair plasma \cite{Tsai-Erber}.
Thus, nonlinear QED effects are already known to be an important
ingredient for pulsar physics. Since the pair plasma gives rise to
radio wave emissions, and because of the large energy scales involved,  
pulsar atmospheres are likely to host other QED effects as well, such
as vacuum nonlinearities in the form of photon--photon scattering. 

An effective theory of photon--photon scattering can be formulated in
terms of the Heisenberg--Euler Lagrangian
\cite{Heisenberg-Euler,Schwinger}. 
In case a dilute plasma is present, the nonlinearities associated
with QED are normally many orders of magnitude smaller than those
due to the classical plasma medium.  
However, here we will show that the QED processes may, for certain
configurations, give a significant contribution  
even in the presence of a plasma. In particular, we will investigate the
propagation of circularly polarized electromagnetic waves in a 
multi-component plasma. The QED effects give rise to
a novel low-frequency mode, which is described by a nonlinear
dispersion relation. The significance of those results are discussed
together with various applications.

According to QED, the non-classical phenomenon of photon--photon scattering
can take place due to the exchange of virtual electron--positron pairs. This
gives rise to vacuum polarization and magnetization currents. In terms of the 
Heisenberg--Euler Lagrangian density \cite{Heisenberg-Euler,Schwinger} we have
\begin{equation}
{\mathscr{L}}_{{\rm EH}}=\epsilon _{0}{\mathscr{F}}+\xi (4{\mathscr{F}}%
^{2}+7{\mathscr{G}}^{2})\ ,  \label{eq:lagrangian}
\end{equation}
where $\xi \equiv 2\alpha ^{2}\epsilon _{0}^{2}\hbar
^{3}/45m_{e}^{4}c^{5}$, ${\mathscr{F}}\equiv \frac{1}{2}(E^{2}-c^{2}B^{2})$, 
${\mathscr{G}}\equiv c{\bf E}\cdot {\bf B}$, $\alpha =
e^2/4\pi\epsilon_0\hbar c$ is the fine structure constant,
$\epsilon_0$ is the vacuum permittivity, $e$ is the electron charge, $c$
the speed of light, $2\pi \hbar $ the Planck constant and $m_{e}$ the
electron mass. The last terms in (\ref{eq:lagrangian}) represent the
effects of vacuum polarization and magnetization, and is the result of
a perturbative expansion of the full Heisenberg--Euler Lagrangian. The
fields are therefore formally constrained by $\omega \ll \omega_e$ and
$|{\bf E}| \ll E_{\text{crit}}$, where $\omega_e = c/\lambda_e =
m_ec^2/\hbar \simeq 8\times 10^{20}$ rad/s and $E_{\text{crit}} =
m_e^2c^3/e\hbar \sim 10^{18}$ V/m, $\lambda_e \simeq 4\times 10^{13}$
m being the Compton wavelength and $E_{\text{crit}}$ the Schwinger
field strength. However, it can be shown that the Lagrangian
(\ref{eq:lagrangian}) is valid upto field strengths surpassing the
Schwinger field, as long as $c|{\bf B}| \geq |{\bf E}|$ is order to
avoid pair creation \cite{Dittrich-Gies,Marklund-Shukla-Eliasson}. 
The QED corrected
Maxwell's vacuum equations take the classical form using 
${\bf D}=\epsilon_{0}{\bf E}+{\bf P}$ and ${\bf H}=({\bf B}/\mu
_{0})-{\bf M}\ ,$ 
where ${\bf P}$ and ${\bf M}$ are of third order in the field amplitudes $%
{\bf E}$ and ${\bf B}$, as shown below in Eqs.\ (\ref{polarization})
and (\ref{magnetization}) respectively, and $\mu _{0}=1/c^{2}\varepsilon _{0}$.

In a medium with polarization ${\bf P}$ and magnetization ${\bf M}$ the
general wave equations for ${\bf E}$ and ${\bf B}$ are 
\begin{subequations}
\begin{equation}
\frac{1}{c^{2}}\frac{\partial ^{2}{\bf E}}{\partial t^{2}}-\nabla ^{2}{\bf E}%
=-\mu _{0}\left[ \frac{\partial ^{2}{\bf P}}{\partial t^{2}}+c^{2}\nabla
(\nabla \cdot {\bf P})+\frac{\partial }{\partial t}(\nabla \times {\bf M)}%
\right]  \label{WaveE}
\end{equation}
and 
\begin{equation}
\frac{1}{c^{2}}\frac{\partial ^{2}{\bf B}}{\partial t^{2}}-\nabla ^{2}{\bf B}%
=\mu _{0}\left[ \nabla \times (\nabla \times {\bf M})+\frac{\partial }{%
\partial t}(\nabla \times {\bf P)}\right] \ .  \label{WaveB}
\end{equation}
\end{subequations}
Furthermore, the effective polarization and magnetization in vacuum due to
photon-photon scattering induced by the exchange of virtual
electron-positron pairs are given by (see, e.g. \cite{Shen-Yu}) 
\begin{subequations}
\begin{equation}
{\bf P}=2\xi \left[ 2(E^{2}-c^{2}B^{2}){\bf E}+7c^{2}({\bf E\cdot B}){\bf B}%
\right] \ ,  \label{polarization}
\end{equation}
and 
\begin{equation}
{\bf M}=-2c^{2}\xi \left[ 2(E^{2}-c^{2}B^{2}){\bf B}-7({\bf E\cdot B}){\bf E}%
\right] \ .  \label{magnetization}
\end{equation}
\end{subequations}

For circularly polarized electromagnetic waves propagating  in a cold
multicomponent plasma rather than in vacuum, the wave operator on the
left-hand sides of Eqs.\ (\ref{WaveE}) and (\ref{WaveB}) is replaced
by 
\begin{equation}
\square \rightarrow \frac{1}{c^{2}}\frac{\partial ^{2}}{\partial t^{2}}
-\nabla ^{2}+\frac{\omega _{p}^{2}}{c^{2}}\rightarrow \frac{-\omega
^{2}+\omega _{p}^{2}}{c^{2}}+k^{2},  \label{substitution}
\end{equation}
where we have assumed that the EM-fields vary as $\exp(ikz-i\omega t)$.
Due to the high symmetry of the circularly polarized EM waves,
most plasma nonlinearities cancel, and the above substitution therefore
holds for arbitrary wave amplitudes, provided that the plasma
frequency depends on the electromagnetic amplitude 
through $\omega_p=\sum_j(n_{0j}q_{j}^{2}/\epsilon_{0}m_{j}\gamma _{j})^{1/2}$,
where the sum is over particle species, $n_{0j}$ denotes the particle
density in the laboratory frame, and the relativistic factor of each
particle species is  
$\gamma_j = (1 + q_j^2 E_0^2/m_j^2 c^2 \omega^2)^{1/2}$,
where $E_0$ denotes the absolute value of the electric field amplitude 
\cite{Stenflo1976,Stenflo-Tsintsadze1979}.

Next, we investigate the regime $\omega^2 \ll k^2 c^2$. From Faraday's
law and the above inequality we note that the dominating QED 
contribution to Eq. (\ref{WaveB}) comes from the term proportional to $B^{2}
{\bf B}$. Combining  Eqs. (\ref{WaveB}) and (\ref{magnetization}), noting 
that $B^2 = B_0^2 = k^2 E_0^2/\omega^2 $\,\, is constant for
circularly polarized EM waves, and using $\omega^2 \ll k^2 c^2$,
i.e. ${\bf M} \simeq 4c^4\xi B^{2}{\bf B}$ and
$|{\bf M}| \gg \omega |{\bf P}|/k$, we thus
obtain from (\ref{WaveB}) the nonlinear dispersion relation 
\begin{equation}
\omega^2 = \frac{2\alpha}{45\pi}\left(
  \frac{E_0}{E_{\text{crit}}} \right)^2 \frac{k^4 c^4}{\omega_p^2+k^2
  c^2} .
  \label{Dispersion-relation} 
\end{equation}
This low-frequency mode makes the particle motion ultra-relativistic
even for rather modest wave amplitudes. For electrons and positrons in
ultra-relativistic motion ($\gamma _{j}\gg 1$) with equal densities $n_{0}$
and elementary charge $\pm e$, we thus use the approximation
$\omega_p^2 \simeq 2en_0 c\omega /\epsilon_0E_0 = 2\omega_{p0}^2
(\omega/\omega_e) (E_{\text{crit}}/E_0)$, where $\omega_{p0} =
(e^2 n_0/\epsilon_0 m_e)^{1/2}$. 
The dispersion relation (\ref{Dispersion-relation}) then reduces to 
\begin{eqnarray}
&& \omega^3 = \frac{\alpha}{45\pi}\left( \frac{\omega_e}{
  \omega_{p0}} \right) \left(
  \frac{E_0}{E_{\text{crit}}} \right)^3 \nonumber \\ 
&& \qquad \times
  \frac{k^4 c^4}{\omega_{p0} + 
  (E_0/E_{\text{crit}})(kc\omega_e /2\omega\omega_{p0})kc} .
\label{Relativistic-DR}
\end{eqnarray}
We note that the ratio $\omega_e/\omega_{p0}$ is much larger than unity for 
virtually all plasmas, i.e.  for electron densities up to $\sim 10^{38}$ 
m$^{-3}$.  In some applications, such as in pulsar astrophysics, it is
convenient  
to re-express the dispersion relation in terms of the
relativistic gamma factor using $E_0/E_{\text{crit}} \simeq
(\omega/\omega_e)\gamma$. Thus, we obtain
\begin{equation}
  \lambda =
  \frac{2}{3}\gamma\lambda_e\sqrt{\frac{2}{5}\pi\alpha}\,\left[
    1 + \sqrt{1 + \frac{4\alpha}{45\pi}\left(
  \frac{\omega_{p0}}{\omega_e} \right)^2\gamma } \right]^{-1/2} 
\label{wavelength}
\end{equation}
from (\ref{Relativistic-DR}) for the wavelength $\lambda =
2\pi/k$. 

Astrophysical environments may exhibit extreme fields. Neutron
stars have surface magnetic field strengths of the order of
$10^{10}-10^{13}\,\mathrm{G}$, while magnetars can reach
$10^{14}-10^{15}\,\mathrm{G}$ \cite{magnetar}, coming close to energy
densities corresponding the Schwinger limit; here, the quantum 
vacuum becomes fully nonlinear. Single-particle QED effects, such as
photon splitting can play a significant role in the understanding and
interpretation of observations from these extreme systems
\cite{Baring-Harding,Harding}.  
In fact, the pair plasma creation in pulsar environments itself rests on
nonlinear QED vacuum effects. The emission of short wavelength photons
due to the acceleration of plasma particles close to the polar caps
results in a production of electrons and positrons as the
photons propagate through the pulsar intense magnetic field 
\cite{magnetosphere}. The precise density of the pair plasma created
in this fashion is difficult the estimate, and the answer is model
dependent. However, given the Goldreich--Julian density $n_{GJ} =
7\times 10^{15} (0.1\,\mathrm{s}/P)(B/10^8\,\mathrm{T})$ m$^{-3}$,
where $P$ is the pulsar period and $B$ the pulsar magnetic field, the
pair plasma density is expected to satisfy $n_0 = M n_{GJ}$, $M$ being
the multiplicity \cite{magnetosphere,Luo-etal}. The multiplicity is
determined by the model through which the pair plasma is assumed to be
created, but a moderate estimate is $M = 10$ \cite{Luo-etal}. Thus,
with these pre-requisites, the density in a hot dense pair plasma is
of the order $10^{18}$ m$^{-3}$, and the pair plasma experiences a
relativistic factor $\sim 10^2 - 10^3$ \cite{Asseo}. Using these
estimates in (\ref{wavelength}), we obtain $\lambda \sim
10^{-12} - 10^{-11}$ m 
\footnote{We note that the pulsar stationary magnetic field and
  relativistic temperature has been neglected in the derivation of
  Eq.\ (\ref{Relativistic-DR}), and the results could therefore be
  modified accordingly.}. On the other hand, the primary beam
will have $n_0 \sim n_{GJ}$ and $\gamma \sim 10^6 -10^7$ \cite{Asseo},
at which (\ref{wavelength}) yields $\lambda \sim 10^{-8} - 10^{-7}$ m.

At the laboratory level, current high laser powers are able to
accelerate particles to highly relativistic speeds. 
Furthermore, pulse self-compression in laser-plasma systems 
may play an important role in attaining power levels well above
current laser limits (see, e.g. Refs.\ \cite{Bulanov-etal,puk03}).
As the intensities approach the Schwinger limit in future laser-plasma
setups, effects of pair-creation and photon--photon scattering have to
be taken into account \cite{Bulanov-etal,Bulanov1}.
Laser-plasma systems can have electron densities of the order $10^{26}$
m$^{-3}$, and laser intensities can be close to
$10^{23}-10^{25}$ W/cm$^2$ \cite{mou98,bob}. Moreover, as stated in
Ref.\ \cite{Bulanov-etal}, laser self-focusing in plasmas could reach
$E_{\text{crit}}$, at which pair creation is likely to follow. In
fact, it has been estimated that the National Ignition Facility
would be able to produce pairs by direct irradiation of a deuterium
pellet \cite{nif}.  On the other hand, the creation of laboratory electron--positron
plasmas is already a feasible task \cite{Surko-etal,Greaves-etal}, as
is the usage of these plasmas for making pair plasma experiments
\cite{Greaves-Surko}. Thus, the possibility to study laser-plasma 
interactions in pair plasmas could be a reality in the nearby future.  
The currently available positron densities in the laboratories are
well below those of regular laser plasma systems, but according to Ref.\
\cite{nif} there is a possibility of reaching densities of order
$10^{27}$ m$^{-3}$. Thus, using $n_0 \sim 10^{26}$ m$^{-3}$, and the
field intensity $10^{16}$ V/m (due to laser self-compression
\cite{Bulanov-etal}) at the wavelength $0.3\times 10^{-6}$ m, we find from
Eq.\ (\ref{Relativistic-DR}) that $\omega \simeq 7.8 \times 10^5$
rad/s, i.e. the frequency is in the LF band. 
 
To summarize, we have presented a new low-frequency circularly polarized 
electromagnetic wave in an electron-positron plasma, taking into account the 
combined action of relativistic charged particle mass variation and
the QED effect  involving photon--photon scattering. The wave frequency strongly 
depends on the wave amplitude. The present mode is shown to play an 
important role in astrophysical settings as well as in the next generation 
laser-plasma systems.

\acknowledgments
This research was partially supported by the Swedish Research Council 
through the contract No. 621-2001-2274 as well as by the Deutsche
Forschungsgemeinschaft through the Sonderforschungsbereich 591 and
the European Commission through the contract No. HPRN-CT-2001-00314.


\end{document}